\documentclass[12pt]{iopart}
\usepackage{graphicx}
\usepackage{iopams}
\begin{document}

\title{Hybrid photonic-bandgap accelerating cavities}

\author{E. Di Gennaro$^1$, C. Zannini$^2$, S. Savo$^2$, A. Andreone$^2$, M. R. Masullo$^3$, G. Castaldi$^4$, I. Gallina$^{4}$, and V. Galdi$^4$}

\address{$^1$ CNISM and Department of Physics, University of Naples ``Federico II,'' Naples, Italy}
\address{$^2$ CNR-INFM ``Coherentia'' and Department of Physics, University of Naples ``Federico II,'' Naples, Italy}
\address{$^3$ INFN -- Naples Unit, Naples, Italy}
\address{$^4$ Waves Group, Department of Engineering, University of Sannio, Benevento, Italy}
\ead{masullo@na.infn.it }

\begin{abstract}
In a recent investigation, we studied two-dimensional point-defected
photonic bandgap cavities composed of dielectric rods arranged
according to various representative periodic and aperiodic lattices,
with special emphasis on possible applications to particle
acceleration (along the longitudinal axis). In this paper, we
present a new study aimed at highlighting the possible advantages of
using {\em hybrid} structures based on the above dielectric
configurations, but featuring {\em metallic} rods in the outermost
regions, for the design of extremely-high quality factor,
bandgap-based, accelerating resonators. In this framework, we
consider diverse configurations, with different (periodic and
aperiodic) lattice geometries, sizes, and dielectric/metal
fractions. Moreover, we also explore possible improvements
attainable via the use of superconducting plates to confine the
electromagnetic field in the longitudinal direction. Results from
our comparative studies, based on numerical full-wave simulations
backed by experimental validations (at room and cryogenic
temperatures) in the microwave region, identify the candidate
parametric configurations capable of yielding the highest quality
factor.
\end{abstract}
\pacs{42.70.Qs, 42.60.Da, 29.20.-c}
\submitto{\NJP}
\maketitle
\section{Introduction and background}
The next-generation colliders (which require large-current and
high-energy beams) and the needs of medical and industrial
applications of accelerators (which ask for compact and
easy-to-fabricate structures) constitute a pressing demand for the
development of resonators based on novel, unconventional concepts.
Amongst a number of problems that the design of new accelerators has
to face, the most important is probably the suppression of
higher-order modes (HOMs) in the resonant accelerating cavities, which
may produce beam instabilities or power losses.

Due to the energy transfer between the bunched beam, which is
traveling through, and the cavity, the amplitude and the distribution
of the electromagnetic (EM) field inside the cavity will be different
from the case in the absence of particles. The harmonic content
of the bunched beam is the driving source of this transfer. Moreover,
the use of high-intensity beams, generally associated with short bunches,
implies an increase of the higher-order harmonics.
The energy transfer between the beam and the cavity will be effective
only if the high-order harmonics are synchronous with the cavity modes.
In this case, a fraction of the excited EM field will remain in the resonant
cavity until it is naturally damped. Such phenomenon, called ``beam loading''
of the accelerating cavity, gives rise to both transverse and longitudinal
coupled-bunch instabilities, and increases linearly with the beam
intensity. As a consequence, the particle current is severely
limited if the instability growth rate is larger than the natural
damping. The fundamental mode can be compensated by varying the
amplitude and phase of the feeding voltage, but the detrimental
HOMs need to be detuned. Possible solutions to this problem (at no
expense to the accelerator performance) are based on connecting
to the cavity a number of waveguides with various cut-off
frequencies. Such HOM-removal mechanism  works very well at
relatively low operational frequencies, but  becomes rather
cumbersome as the frequency increases.

Within this context, it is presently a real challenge to design and
build a compact, HOM-free, accelerating structure, able at the same
time to efficiently couple the EM field to the particle beam.

Periodic photonic crystals (PCs) have been proposed in the past as
candidates for accelerating cells in microwave or laser-driven
particle accelerators \cite{Cowan}. The key concept underlying these
structures is the presence of a photonic bandgap (PBG),
typical of dispersive media, which prevents, within a specific
frequency band, the EM propagation along the periodicity directions
of the PC. In such structures, a PBG cavity (or waveguide) can be
realized by introducing one or more localized lattice defects,
thereby producing ``field trapping'' nearby the defect zone within
the forbidden frequency window. This mechanism therefore allows to
design a frequency-selective structure for the EM propagation,
acting, e.g., as a perfectly ``reflecting wall'' at a certain
frequency, while exhibiting a transparent response in the remaining
part of the spectrum. By adequately shaping the geometry around the
defect, the trapped mode can be optimized so as to work as the
operating accelerating mode. Metallic PC structures have already
been used to realize a new kind of large-gradient accelerator with
an effective suppression of HOM wakefields \cite{Smirnova}.
Prototypes of fully-metallic (super- and normal-conducting)
mono-modal PC cavities have been constructed and tested at different
working frequencies \cite{Masullo}.

Previous (numerical and experimental) studies have also demonstrated
that {\em all-dielectric} structures can be used to confine the desired
excited mode, eliminating or reducing the characteristic metallic
losses at the frequency of operation. Dielectric materials may also
be beneficial in order to cope with radio-frequency (RF) breakdown
phenomena that may easily occur in structures where large
accelerating field gradients need to be achieved.

Moreover, the existence of PBG and related phenomena is not
restricted to {\em periodic} crystals. In fact, a large body of
numerical and experimental studies have demonstrated the possibility
of obtaining similar effects also in {\em aperiodically-ordered}
structures, typically referred to as ``photonic quasicrystals''
(PQCs) (see, e.g., \cite{Steurer,DellaVilla,Lavrinenko} for recent
reviews of the subject). PQC structures are receiving a growing
attention in a variety of fields and application scenarios (see,
e.g., \cite{Macia}), mainly driven by the higher level of (weak)
symmetry achievable and the extra degrees of freedom available, as
compared to their periodic counterparts. In particular, experimental
studies on PQC-based optical microcavities have demonstrated the
possibility of obtaining  high quality factors and small modal
volumes \cite{Nozaki}, thereby providing further degrees of freedom
in tailoring the mode confinement. In this framework, it is also
worth mentioning possible alternative strategies based on
brute-force numerical optimization of the spatial lattice geometry
\cite{Bauer}. For the case of dielectric structures, by comparison
with a periodic lattice exhibiting comparable performance, this
strategy may allow a considerable saving in the number of rods,
although it is not entirely clear how this sparsification translates
into a reduction of the overall structure size.

In a recent investigation \cite{DiGennaro}, we explored the possible
application of 2-D point-defected PQC cavities composed of
dielectric rods arranged according to various representative
aperiodic (Penrose and dodecagonal) geometries, and terminated by
two metallic plates in order to confine the field along the
longitudinal direction. More specifically, we carried out a
parametric study of the confinement properties as a function of the
structure size, filling fraction, and losses, so as to identify the
best performing configurations, and we compared them with a
reference periodic counterpart. Although the results were very
encouraging, we found that the major limitation in the development
of compact, intrinsically mono-modal (and hence HOM-free),
high-quality-factor cavities arises from the conduction losses due
to the metallic plates, irrespective of the geometry considered. A
possible device to overcome this limitation amounts to replacing
copper with a superconducting material, thereby rendering the
in-plane radiation the dominant loss mechanism. In this framework, a
systematic study needs to be pursued in order to assess to what
extent the actual improvements in the cavity performance might
justify the higher complexity introduced by the necessity of
operating at cryogenic temperatures.

This paper, following up on our previous work, is aimed at
highlighting the possible advantages of using {\em metallic} rods in
the outermost regions of (periodic and aperiodic) dielectric
structures, for the design and fabrication of PBG-based {\em hybrid}
(normal metallo-dielectric and superconducting metallo-dielectric)
high-quality-factor accelerating resonators. The basic underlying
idea is to find out a suitable trade-off between dielectric and
metal content, so as to improve the in-plane confinement without
significantly increasing the conduction losses. To this aim, we
compare different configurations of these hybrid structures, with
diverse sizes, lattice geometries, and dielectric/metal fractions.

Accordingly, the rest of the paper is laid out as follows. In Sec.
\ref{Sec:Numerical}, we present the results of our numerical
full-wave studies, first on dielectric-rod structures (focusing on
the effects of the  metallic plates), and subsequently on {\em
hybrid} structures consisting of dielectric and metallic rods. In
Sec. \ref{Sec:Exp}, we validate the above results via experimental
measurements at room and cryogenic temperatures. Finally, in Sec.
\ref{Sec:Concl}, we provide some concluding remarks and hints for
future research.

\section{Numerical studies}
\label{Sec:Numerical}
Our numerical full-wave studies of the EM response of the structures
of interest are based on the combined use of the 3-D commercial
software CST Microwave Studio \cite{CST} (for modeling volumetric
and surface losses) and an in-house 2-D simulator based on the
finite-difference-time-domain (FDTD) technique \cite{Taflove} (for
modeling the radiative losses not accounted for in the 3-D
simulator).

\subsection{Dielectric PBG cavities}
\begin{figure}[hbtp!]
\centering
\includegraphics[width=0.6 \textwidth]{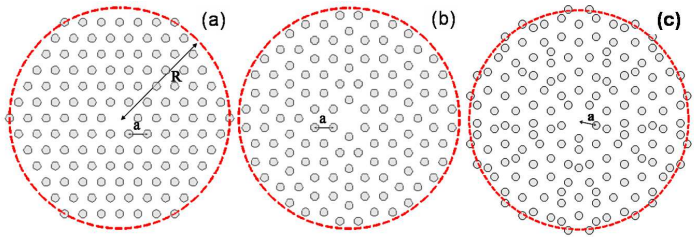}\\
\caption{Point-defected PBG mono-modal cavities, with periodic triangular (a), and
aperiodic dodecagonal (b) and Penrose (c) lattice geometries.
}\label{geom}
\end{figure}

The dielectric structures of interest are composed of sapphire
cylindrical rods of radius $r$ and relative dielectric permittivity
9.2, with typical loss-tangent values ranging between $10^{-6}$ and
$10^{-8}$, depending on the temperature of operation. As in
\cite{DiGennaro}, the rods are arranged according to two
representative PQC geometries, based on the dodecagonal
(12fold-symmetric \cite{Oxborrow}, \Fref{geom}(b)) and Penrose
(10fold-symmetric \cite{Senechal}, \Fref{geom}(c)) tilings,
respectively, and a reference periodic (triangular) PC structure
(\Fref{geom}(a)). The removal of the central rod creates the defect
region and allows for the beam transit aperture. The mode of
interest for the particle acceleration along the longitudinal
direction is the $TM_{010}-like$ (fundamental mode), with
the electric field parallel to the rods.

All configurations are characterized by a lattice constant
(corresponding to the period in the triangular case, or to the tile
sidelength in the aperiodic cases, cf. \Fref{geom}) chosen as $a=
0.75$ cm, so as to yield comparable values of the fundamental mode
resonant frequency (around 16.5 GHz). The rods height is $h= 0.6$
cm. Different transverse sizes are considered, by varying the radius
$R$ as an integer multiple of the lattice constant $a$ (see
\Fref{geom}). As a figure of merit, we use the standard
quality factor,
\begin{equation}
Q=\frac{\omega_{0}{\cal E}}{{\cal P}},
\label{eq:Q}
\end{equation}
where $\omega_0$ is the resonant radian frequency, ${\cal E}$ is the
EM energy stored in the cavity, and ${\cal P}$ is the average power
loss. In our simulations, the resonant frequency and the quality
factors pertaining to volumetric and surface losses are computed via
standard post-processing routines available in the CST Microwave
Studio eigen-solver \cite{CST}. The radiative quality factor is
instead computed from the 2-D FTDT analysis (with all dielectric and
metallic elements assumed as lossless), by processing the
time-signal evolution via a harmonic inversion tool \cite{harminv}
based on a low-storage ``filter diagonalization method.'' The
overall quality factor $Q_T$ can then be obtained by combining the
conducting ($Q_C$), dielectric ($Q_D$), and radiative ($Q_R$)
quality factors via
\begin{equation}
\frac{1}{Q_T}=\frac{1}{Q_{C}}+\frac{1}{Q_{R}}+\frac{1}{Q_{D}}.
\label{eq:QT}
\end{equation}
Note that, within the parametric ranges of interest, the dielectric
quality factor $Q_D$ is much higher than the other two factors, and
its contribution in (\ref{eq:QT}) is accordingly negligible.

As shown in \cite{DiGennaro}, and compactly summarized
in \Tref{qfac}, for a moderate cavity size ($R\leq 5 a$) and a
filling factor $r/a = 0.2$, a judicious choice of a PQC geometry
turns out to provide sensible improvement in the field confinement
as compared with the periodic reference configuration.
\begin{table}[hbtp!] \caption{\label{qfac}Simulation results for the
selected lattice geometries, assuming cavity sizes $R=3a, 4a, 5a$
and filling factor $r/a=0.2$. $Q_C$, $Q_R$, and $Q_T$ are the
conducting, radiative, and total quality factors, respectively. The
last column indicates the weight of the conduction losses in the
total quality factor.}
\begin{indented}
\item[]\begin{tabular}{@{}lcccc}
\br

$R=3a$& $Q_C$&  $Q_R$&   $Q_T$&    ($1 - \frac{Q_T}{Q_R}$)\\
\mr
Triangular&   $1.05\times 10^4$& $  7.74\times 10^2$& $  7.20\times 10^2$&  0.05\\
Dodecagonal&    $1.07\times 10^4$&$   1.87\times 10^3$& $  1.65\times 10^3$&  0.09\\
Penrose&    $1.06\times 10^4$& $  4.37\times 10^2$&  $ 4.23\times 10^2$&  0.02\\
\mr $R=4a$&&&&\\
\mr
Triangular&   $1.19\times 10^4$&   $5.70\times 10^3$&  $ 3.80\times 10^3$& 0.32\\
Dodecagonal&   $ 1.18\times 10^4$&  $ 1.72\times 10^4$&  $ 7.00\times 10^3$&    0.59\\
Penrose&    $1.14\times 10^4$&   $5.00\times 10^3$&  $ 3.48\times 10^3$& 0.30\\
\mr $R=5a$&&&&\\
\mr
Triangular&   $1.19\times 10^4$&  $ 4.08\times 10^4$&  $ 9.20\times 10^3$& 0.77\\
Dodecagonal&  $  1.19\times 10^4$&  $ 9.30\times 10^4$&  $ 1.05\times 10^4$&    0.89\\
Penrose&    $1.22\times 10^4$&  $ 1.47\times 10^5$&  $ 1.13\times 10^4$& 0.92\\
\br

\end{tabular}
\end{indented}
\end{table}

In dielectric-rod cavities, as expectable, the conducting quality
factor $Q_C$ is almost the same for all configurations and cavity
sizes, since it depends mainly on the surface conductivity of the metallic plates.

Conversely, the radiative quality factor $Q_R$ strongly depends on the geometry and size of each structure. The field confinement in these cavities is
weaker than that achievable via fully-metallic structures, and represents the main factor affecting the cavity performance when very compact ($R=3a$)
structures are needed.

For $R=5a$, instead, $Q_R$ improves from $4.08\times 10^4$
(triangular) to $1.47\times 10^5$ (dodecagonal). For this size, the
total quality factor $Q_T$ of dielectric (periodic or aperiodic)
structures is much higher than those obtained in the case of
fully-metallic periodic PBG cavities ($\sim 4 \times 10^3$ at room
temperature) of comparable resonant frequency \cite{Masullo2}. This
is mainly due to the reduction of conduction losses resulting from
the use of dielectric (instead of metallic) rods.

In such a case, therefore, the primary source of dissipation is
given by the surface losses of the metallic plates. The replacement
of copper with a superconducting material appears the way to go,
even if the required low-temperature operation implies an increased
complexity. Nevertheless, in order to achieve the performance
required for accelerating cavities, one still needs to reduce the
radiation leaks, which would otherwise limit the total quality
factor.

\subsection{Hybrid PBG cavities}
In order to reduce the radiative losses, without sacrificing the
performance improvement attainable via superconducting technologies,
one may think of replacing {\em some} dielectric rods (intuitively,
those located in the outermost regions, where the field is weaker),
with metallic ones, thereby obtaining a ``hybrid''
metallo-dielectric PBG cavity. We present here the results
pertaining to hybrid structures of size $R=5a$, based on the above
selected lattice geometries (triangular, dodecagonal, Penrose).

\Tref{qhyb} compares the FDTD-simulated radiative quality factors
pertaining to the dielectric-rod reference case with those
pertaining to hybrid PBG cavities featuring one or two peripheral
``rings'' made of metallic (copper) rods. Here, and henceforth, the
hybrid structures are labeled as $D+M$, with $D$ and $M$ denoting
the number of rings made of dielectric and metallic rods,
respectively. Note that, in the aperiodic cases, the ``rings'' are
not regularly shaped, and their definition may be ambiguous. In our
simulations, they were defined via radial inequalities (e.g., for a
total cavity size of $R=5a$, the outermost ring is defined as
exterior to the radial domain $R'=4a$, and so on); this ensures the
inclusion of a comparable number of metallic rods for the different
lattice geometries.

\begin{table}[hbtp!]
\caption{\label{qhyb}Simulated radiative quality factors $Q_R$ for
hybrid PBG cavities of total size $R=5a$ (with $r/a=0.2$), featuring
zero (i.e., fully-dielectric), one, and two peripheral rings of
metallic (copper) rods.}

\begin{indented}
\item[]\begin{tabular}{@{}cccc}
\br

Dielectric+Metallic&   Triangular&   Dodecagonal&   Penrose\\
\mr
5 + 0&   $4.08\times 10^4$&   $9.30\times 10^4$&  $1.47\times 10^5$\\
4 + 1&   $1.78 \times 10^6$&   $7.44 \times 10^6$&  $2.51 \times 10^6$\\
3 + 2&   $2.50 \times 10^8$&   $5.85 \times 10^7$&  $3.93 \times 10^8$\\
\br

\end{tabular}
\end{indented}
\end{table}

One readily observes that the inclusion of metallic rods
dramatically improves the confinement properties of the PBG
cavities. In particular, in the periodic case, there is a
two-order-of-magnitude step increase in the radiative quality
factor, which brings its value to over $10^8$ (very close to what
predicted for the Penrose geometry) when two peripheral copper rings
are included. The dodecagonal geometry, which exhibits the best
performance in the 4+1 configuration, is outperformed by the other
geometries in the 3+2 configuration.

\begin{figure}[hbtp!]
\centering
\includegraphics[width=0.95\textwidth]{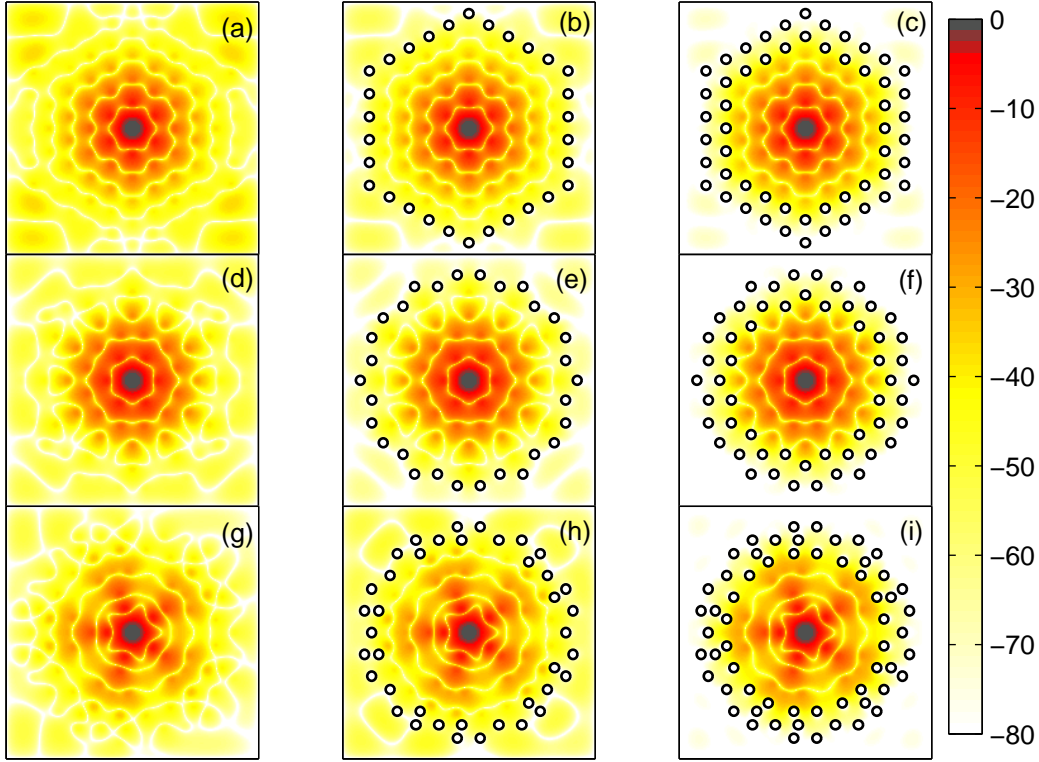}\\
\caption{Simulated electric-field intensity maps (in dB) for the hybrid PBG
cavities of size $R=5a$ (with $r/a=0.2$), featuring zero, one or two peripheral ``rings'' of copper rods (displayed as black empty circles), and different lattice geometries:
triangular ((a), (b), (c)), dodecagonal ((d), (e), (f)), Penrose ((g), (h), (i)).
}\label{hybrcomp}
\end{figure}

The field-confinement improvements are also evident in the
corresponding (transverse) electric-field maps shown in
\Fref{hybrcomp}.

Specifically, Figs. \ref{hybrcomp}(a), \ref{hybrcomp}(d), and
\ref{hybrcomp}(g) show the results pertaining to the dielectric-rod
cavities (5+0 configuration). The field maximum intensity (centered
at the defect position) is almost the same in the three different
cases, but the spatial distribution evidences the better confinement
properties of the Penrose geometry, as confirmed by the data
reported in \Tref{qfac}. The improvement in the radiative quality
factor of the hybrid cavities is already sensible when the first
(outermost) peripheral metallic ring is included (see Figs.
\ref{hybrcomp}(b), \ref{hybrcomp}(e), and \ref{hybrcomp}(h)), and
becomes striking in the 3+2 configuration (see Figs.
\ref{hybrcomp}(c), \ref{hybrcomp}(f), and \ref{hybrcomp}(i)). This
is in agreement with the trend shown in \Tref{qhyb}. More difficult,
however, is to discern from the plots the different performance in
terms of $Q_R$ values exhibited by the three geometries for each
hybrid (4+1 or 3+2) configuration.

\begin{figure}[hbtp!]
\centering
\includegraphics[width=0.65\textwidth]{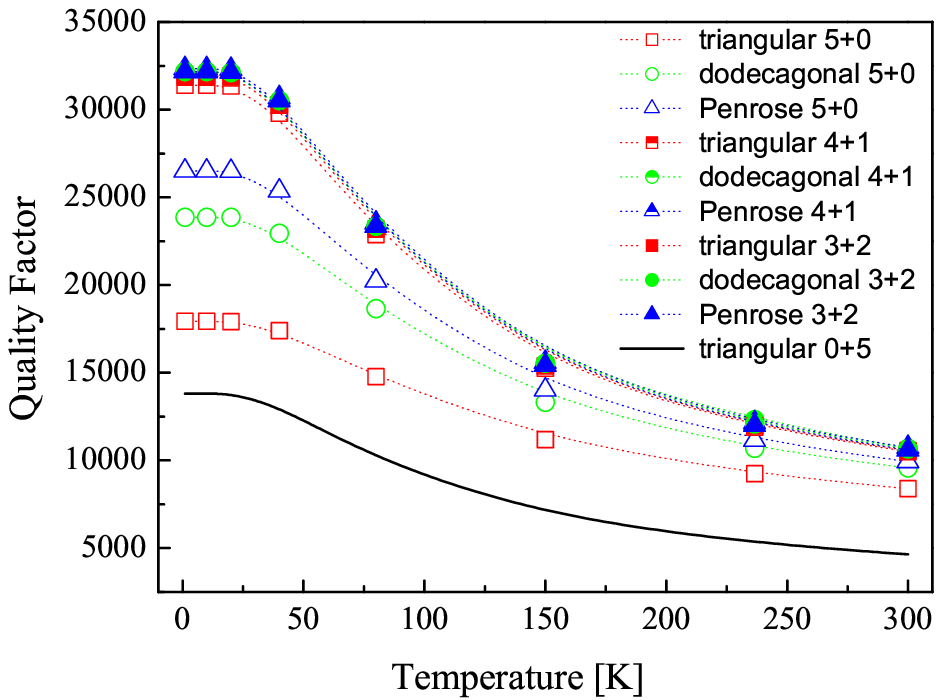}\\
\caption{Simulated temperature dependence of the total quality
factor for triangular, dodecagonal and Penrose PBG hybrid cavities
of size $R = 5a$ (with $r/a=0.2$), and featuring zero, one or two
copper rings, compared with the behavior expected for a fully
metallic triangular cavity (solid curve). }\label{Sim_T}
\end{figure}

\Fref{Sim_T} shows the simulated total quality factors pertaining
the three lattice geometries, as a function of the temperature $T$
and of the number of metallic rings. Direct-current conductivity is
assumed to vary (with the temperature) according to the data
reported in \cite{Inagaki} for high-purity electropolished copper.
As a reference, the behavior of a fully-metallic periodic structure
is also displayed.

Looking at these behaviors, the advantage of using a dielectric-rod
structure (empty markers) instead of a fully-metallic one (solid
line) is fairly clear. Similarly, hybrid structures with one
(semi-empty markers) or two (full markers) metallic rings outperform
(the more the lower the temperature) the fully-dielectric ones. It
is also evident that the inclusion of metallic rings progressively
brings the radiation losses to a negligible level; in this regime,
the dominant source of dissipation comes from the metallic plates,
and the responses (as a function of the temperature) of the
different geometries tend to become identical. A possible solution
might be the replacement of copper with a high-temperature
superconductor (HTS), to cover the inner surface of the confinement
plates. Setting the operational temperature at about 30 K, this
would determine a three-order-of-magnitude reduction of the surface
losses, and therefore a corresponding increase of the related
conduction quality factor. It is important to stress that use of HTS
peripheral rod rings would only add further complexity to the
structure (because of the necessity of efficiently cooling down them
too), without any significant improvement in the overall quality
factor.

\begin{figure}[hbtp!]
\centering
\includegraphics[width=0.65\textwidth]{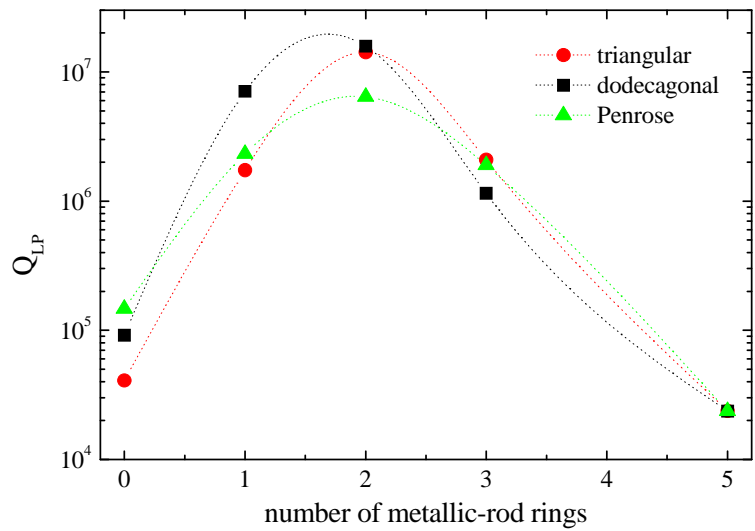}\\
\caption{Estimated total quality factor $Q_{LP}$ as a function of
the number of metallic (copper) rod rings for triangular,
dodecagonal, and Penrose PBG hybrid structures of size $R = 5a$
(with $r/a=0.2$) at a temperature of 30 K, assuming lossless plates
and dielectric loss-tangent of $10^{-8}$. Dotted curves are
guide-to-eye only. }\label{Qvsrods}
\end{figure}

It therefore appears that a judicious combination of {\em i)}
superconducting plates, {\em ii)} low-loss dielectric rods (in the
interior region), and {\em iii)} metallic rods (in the outermost
region) may open up new perspectives in the development of novel
monomodal, PBG based, high-quality-factor open cavity for the
acceleration of energetic particle beams at very high operational
frequencies. The use of peripheral metallic rings certainly improves
the confinement properties of the PBG resonators, while maintaining
the advantages foreseen for dielectric cavities (reduction of
breakdown phenomena, moderate fabrication complexity, etc.).
\Fref{Qvsrods} displays the total quality factor $Q_{LP}$ of a
hybrid PBG cavity of size $R=5a$ as a function of the number of
metallic rings for the geometries of interest, at a temperature of
operation of 30 K, assuming lossless plates and a dielectric
loss-tangent of $10^{-8}$ (which is a reasonable value for
single-crystal sapphire at low temperatures), so as to better
highlight the role of the copper rods.

As can be observed, $Q_{LP}$ first increases (reaching its maximum
for the 3+2 configuration), and then it starts decreasing (reaching
its minimum for the fully-metallic 0+5 configuration). As expected,
increasing the number of metallic rings {\em levels} the performance
of the hybrid cavities, irrespective of the different spatial
arrangements of the rods. Another interesting feature observable in
\Fref{Qvsrods} is that the Penrose geometry, in spite of its higher
radiative quality factor, is largely outperformed by the other two
geometries in the 3+2 configuration. This is attributable to the
slight ($\sim$10\%) larger number of copper rods in outermost
metallic rings (as compared to the triangular and dodecagonal
cases), with a detrimental effect on the level of conduction losses
for this geometry.

\section{Experimental results}
\label{Sec:Exp}

\begin{figure}[hbtp!]
\centering
\includegraphics[width=0.65\textwidth]{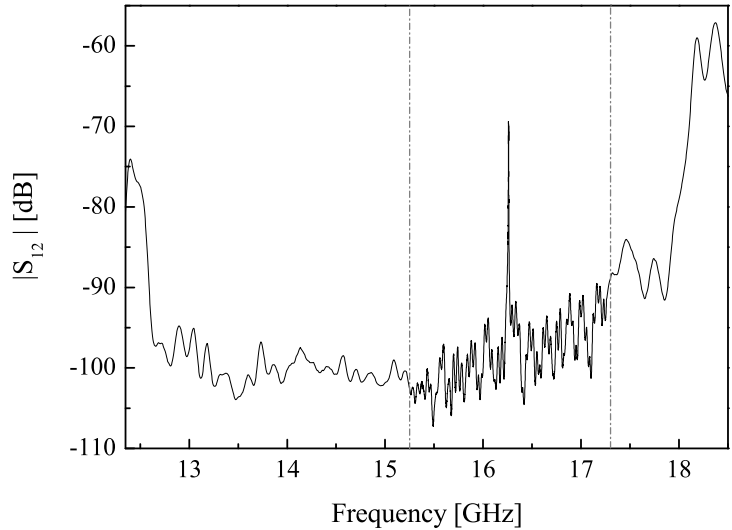}\\
\caption{Transmission parameter $|S_{12}|$ measured as a function of
the frequency for a Penrose (4+1) point-defect cavity. In the
frequency region between the dashed-dotted vertical line, the
frequency sampling has been increased in order to capture the sharp
resonant peak.}\label{scatparam}
\end{figure}

In order to validate the above findings and explore their
technological viability, we fabricated some prototypes of the
simulated structures, by suitably placing the dielectric
(single-crystal sapphire) and metallic (oxygen-free
high-conductivity copper) rods (of radius $r = 0.15$ cm and height
$h = 0.6$ cm) between two copper plates, inserted in a cryogenic
box. The whole system is then cooled down to about 100 K by
introducing liquid nitrogen. The operation temperature is monitored
using a Si-diode sensor placed on one of the copper plates. Both the
feed and pick-up antennas are placed on the top plate, far enough
from the central region where the EM field reaches its maximum to
ensure that all measurements are performed in the weak coupling
limit. The resonant cavity is then connected to a HP8720C network
analyzer and the quality factors at room and cryogenic temperatures
are evaluated by the standard $-3$ dB method, looking at the
frequency transmission curve at the resonance. For the experimental
characterization, we considered hybrid cavities with the three
different geometries and $R=5a$, having a single peripheral metallic
ring (4+1 configuration). As a representative example,
\Fref{scatparam} shows the measured transmission parameter
$|S_{12}|$ pertaining to the Penrose geometry, from which the
\textit{monomodality} of the cavity within the whole bandgap
frequency region (12.5--18 GHz) is fairly evident.

In \Tref{Table3}, the experimental quality factors are reported at
300 K and 100 K, and compared with the results obtained from the
numerical simulations. Note that, unlike in \cite{Inagaki}, the
surface of the copper used in our experiments was not chemically
treated or polished (to remove impurities, oxide layers, etc.). This
was accordingly taken into account in the numerical simulations by
slightly (10 \%) increasing the value of the copper surface
impedance with respect to the data reported in \cite{Inagaki}.

\begin{table}[hbtp!]
\caption{\label{expres} Measured total quality factor at room (300
K) and cryogenic (100 K) temperatures for triangular, dodecagonal
and Penrose PBG hybrid cavities with $R = 5a$ and a single
peripheral metallic ring (i.e., 4+1 configuration).}
\begin{indented}
\item[]\begin{tabular}{@{}lcc|cc}
\br

Geometry& $Q_{exp}^{100K}$&  $Q_{sim}^{100K}$&       $Q_{exp}^{300K}$&  $Q_{sim}^{300K}$\\
\mr
Triangular&   $1.84\times 10^4$&  $ 1.90\times 10^4$&  $ 1.14\times 10^4$& $ 1.05\times 10^4$\\
Dodecagonal&  $  1.95\times 10^4$&  $ 1.94\times 10^4$&  $ 1.12\times 10^4$&$ 1.07\times 10^4$\\
Penrose&    $2.00\times 10^4$&  $ 1.96\times 10^4$&  $ 1.05\times 10^4$&$ 1.07\times 10^4$\\
\br

\end{tabular}
\end{indented}
\label{Table3}
\end{table}

There is a very nice agreement between measurements and simulations,
confirming the validity of our initial assumptions. The data show
that losses, at both room and cryogenic temperatures, are
essentially dominated by the conductive contribution due to the
metallic plates, and consequently determine an upper limit for the
total quality factor of the order of $10^4$, irrespective of the
geometrical configurations, as already evidenced in \Fref{Sim_T}.
One can accordingly conjecture that the insertion of superconducting
plates, with the corresponding reduction of conduction losses of
three or more orders of magnitude, would have already for this
configuration a tremendous impact on the quality factor (and hence
on the overall performance) of an accelerating cavity operating at
such high frequencies. Another interesting conclusion that can be
drawn is that the use of conventional low critical temperature
materials like niobium, very common in the development of
superconducting accelerating cavities, in this case is unnecessary,
since the overall quality factor of hybrid cavities of such compact
size would be inherently limited (by radiation losses) to values on
the order of $10^7$ (see \Fref{Qvsrods}). HTS materials may be used
instead, with an obvious simplification of the related cryogenic
technology, and corresponding cost reduction.

\section{Conclusions}
\label{Sec:Concl}

In this paper, we have explored hybrid configurations of
point-defected PBG cavities, showing that a clever blend of
superconducting materials, low-loss dielectrics, and  highly
conducting metals may pave the way to the development of novel
monomodal, compact, high-performance, accelerating cavities. Via a
systematic study of geometrical configurations, size, and
dielectric/metal fractions, we showed that suitably dimensioned
hybrid open structures may attain high in-plane EM radiation
confinement, without significant increase in the conduction losses.
The exploitation of superconducting materials (in the terminating
plates) would render the fabrication of this new type of resonators
extremely rewarding, even if at the expense of a higher operational
complexity introduced by the cryogenic technology. Our preliminary
experimental results at 100 K show that this route is
technologically viable, especially for the development of very
compact, hybrid, PBG cavities based on HTS materials.

\ack This work was supported in part by the Campania Regional Government via a
2006 grant (L.R. N. 5 - 28.03.2002) on ``Electromagnetic-bandgap
quasicrystals: Study, characterization, and applications in the
microwave region.'' Stimulating discussions with Prof.
V. G. Vaccaro (University of Naples ``Federico II''), as well as the technical support of
F. M. Taurino and S. Marrazzo, are gratefully acknowledged.

\section*{References}
\bibliography{DiGennaro_NJP_biblio_revised}

\begin{thebibliography}{10}

\bibitem{Cowan}
B.~Cowan, M.~Javanmard, and R.~Siemann.
\newblock Photonic crystal laser accelerator structures.
\newblock In {\em Proc. Particle Accelerator Conference}, volume~3, pages
  1855--1857, Portland, OR, USA, 2003.

\bibitem{Smirnova}
E.~I. Smirnova, A.~S. Kesar, I.~Mastovsky, M.~A. Shapiro, and R.~J. Temkin.
\newblock Demonstration of a 17-{GHz}, high-gradient accelerator with a
  photonic-band-gap structure.
\newblock {\em Phys. Rev. Lett.}, 95(7):074801, 2005.

\bibitem{Masullo}
M.~Masullo, A.~Andreone, E.~Di Gennaro, F.~Francomacaro, G.~Lamura,
  V.~Palmieri, D.~Tonini, M.~Panniello, and V.~Vaccaro.
\newblock {PBG} superconducting resonant structures.
\newblock In {\em Proc. European Particle Accelerator Conf.}, pages 454–--456,
  Edinburgh, Scotland, 2006.

\bibitem{Steurer}
W.~Steurer and D.~Sutter-Widmer.
\newblock Photonic and phononic quasicrystals.
\newblock {\em J. Phys. D}, 40(13):R229--R247, 2007.

\bibitem{DellaVilla}
A.~Della Villa, V.~Galdi, S.~Enoch, G.~Tayeb, and F.~Capolino.
\newblock Photonic quasicrystals: Basics and examples.
\newblock In F.~Capolino, editor, {\em Metamaterials Handbook, vol. I},
  chapter~27. CRC Press, Boca Raton, FL, USA, 2009.

\bibitem{Lavrinenko}
D.~N. Chigrin and A.~V. Lavrinenko.
\newblock Photonic applications of two-dimensional quasicrystals.
\newblock In F.~Capolino, editor, {\em Metamaterials Handbook, vol. II},
  chapter~28. CRC Press, Boca Raton, FL, USA, 2009.

\bibitem{Macia}
E.~Maci\'{a}.
\newblock The role of aperiodic order in science and technology.
\newblock {\em Rep. Progr. Phys.}, 69(2):397--441, 2006.

\bibitem{Nozaki}
K.~Nozaki and T.~Baba.
\newblock Quasiperiodic photonic crystal microcavity lasers.
\newblock {\em Appl. Phys. Lett.}, 84(24):4875--4877, 2004.

\bibitem{Bauer}
C.~A Bauer, G.~R. Werner, and J.~R. Cary.
\newblock Truncated photonic crystal cavities with optimized mode confinement.
\newblock {\em J. Appl. Phys.}, 104:053107, 2008.

\bibitem{DiGennaro}
E.~Di Gennaro, S.~Savo, A.~Andreone, V.~Galdi, G.~Castaldi, V.~Pierro, and
  M.~R. Masullo.
\newblock Mode confinement in photonic quasicrystal point-defect cavities for
  particle accelerators.
\newblock {\em Appl. Phys. Lett.}, 93(16):164102, 2008.

\bibitem{CST}
{\em {CST Microwave Studio}}.
\newblock {CST -- Computer Simulation Technology}, Wellesley Hills, MA, USA,
  2008.

\bibitem{Taflove}
A.~Taflove and S.~C. Hagness.
\newblock {\em Computational Electrodynamics: The Finite-Difference Time-Domain
  Method, Third Edition}.
\newblock {Artech House}, Norwood, MA, USA, 2005.

\bibitem{Oxborrow}
M.~Oxborrow and C.~L. Henley.
\newblock Random square-triangle tilings: {A} model for twelvefold-symmetric
  quasicrystals.
\newblock {\em Phys. Rev. B}, 48(10):6966--6998, 1993.

\bibitem{Senechal}
M.~Senechal.
\newblock {\em Quasicrystals and Geometry}.
\newblock Cambridge University Press, Cambridge, UK, 1995.

\bibitem{harminv}
Harminv.
\newblock http://ab-initio.mit.edu/harminv.

\bibitem{Masullo2}
M.~R. Masullo, A.~Andreone, E.~Di~Gennaro, F.~Francomacaro, G.~Lamura, V.G.
  Vaccaro, G.~Keppel, V.~Palmieri, and D.~Tonini.
\newblock A study on a mono-modal accelerating cavity based on photonic band
  gap concepts.
\newblock In {\em Proc. Int. Workshop on "Physics at a multi-MW proton
  source"}, pages 15--19, Firenze, Italy, 2004.

\bibitem{Inagaki}
S.~Inagaki, E.~Ezura, J.F. Liu, and H.~Nakanishi.
\newblock Thermal expansion and microwave surface reactance of copper from the
  normal to anomalous skin effect region.
\newblock {\em J. Appl. Phys.}, 82(11):5401--5410, 1997.

\end{thebibliography}
\bibliographystyle{unsrt}
\end{document}